\documentclass[runningheads]{svmult}

\usepackage{makeidx}   
\usepackage{graphicx}  
\usepackage{subeqnar}  
\usepackage{multicol}  
\usepackage{physprbb}  
\makeindex             

\begin{document}
\title*{Red Giants Survey in $\omega$~Cen: preliminary FLAMES GTO results}
\toctitle{Metal rich stars in $\omega$~Cen:
preliminary FLAMES GTO results}
%
%
\titlerunning{Metal rich stars in $\omega$~Cen}
%
\author{Elena Pancino\inst{1}}
\authorrunning{Elena Pancino}
%
%
\institute{INAF - Osservatorio Astronomico di Bologna\\
     via Ranzani 1, I-40127 Bologna, Italy}

\maketitle              

\begin{abstract}

I present preliminary results for a sample of $\sim$700 red giants in
$\omega$~Cen, observed during the Ital-FLAMES Consortium GTO time in May 2003,
for the Bologna Project on $\omega$~Cen. Preliminary Fe and Ca abundances
confirm previous results: while the metal-poor and intermediate populations
show a normal halo $\alpha$-enhancement of [$\alpha$/Fe]$\simeq$+0.3, the most
metal-rich stars show a significantly lower [$\alpha$/Fe]$\simeq$+0.1. If the
metal-rich stars have evolved within the cluster in a process of
self-enrichment, the only way to lower their $\alpha$-enhancement would be SNe
type Ia intervention. 

\end{abstract}

\section{Introduction}

The treatment of the large data volumes that can be obtained with large
telescopes and multi-objects spectrographs, such as FLAMES at the ESO VLT,
requires a major upgrade of the methods commonly used to reduce and analyze
high and medium resolution spectra. 

We are therefore developing a set of routines for an automatic or
semi-automatic abundance analysis of stellar spectra based on equivalent widths
(EW). The first product is DAOSPEC, a code developed by P. B. Stetson for
automatic EW measurement (http://cadcwww.hia.nrc. ca/stetson/daospec/). The
preliminary abundance analysis presented here is the first step of an
iteractive and automatic procedure under development at the Bologna
Observatory.

\section{Analysis and Results}

Part of the Ital-FLAMES Guranteed Time has been devoted to the study of
$\omega$~Cen, within the framework of the Bologna Project on $\omega$~Cen
\cite{pancino:bokey}. 700 giants have been observed with the ESO-VLT GIRAFFE
spectrograph in May 2003, with a resolution of R$\sim$20,000 and a
S/N$\simeq$50--150 per pixel, spanning more than 4 magnitudes and covering all
known sub-populations in the cluster. The spectral ranges observed (Setups
HR09, HR13, HR11 and HR15) allow the determination of iron peak, $\alpha$,
$s$-process elements and of Eu.

The spectra have been reduced with the GIRAFFE BLDRS pipeline developed at the
Geneva Observatory. EW have been measured with DAOSPEC \cite{pancino:ps05},
based on a linelist produced with the Vienna Atomic Line Database (VALD)
\cite{pancino:vald}. Preliminary estimates of the stellar parameters $T_{eff}$,
$\log g$, $v_t$ and [M/H] have been obtained from the WFI photometry published
by \cite{pancino:p00} and the color-temperature calibration by
\cite{pancino:a99}. MARCS model stellar atmospheres \cite{pancino:edv} have
been employed together with the abundance analysis code by \cite{pancino:s67}
to produce a first guess abundance of Fe and Ca, strictly based on the
photometric parameter estimates.

\begin{figure}[t]
\begin{center}
\includegraphics[angle=270,width=.55\textwidth]{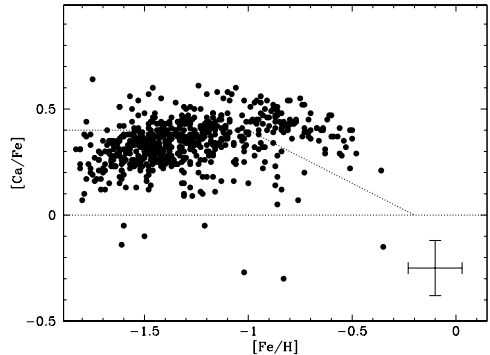}
\end{center}
\caption[]{[Ca/Fe] versus [Fe/H] for the $\sim$700 stars observed with GIRAFFE.
Typical errorbars are plotted in the lower-right corner. The typical disk-halo
behaviour is represented by dotted lines.}
\label{pancino:cafe}
\end{figure}

The resulting [Ca/Fe] versus [Fe/H] plot is shown in Fig.~\ref{pancino:cafe},
where except for a few outliers that will have to be manually inspected, a
clear trend appears: [Ca/Fe] slowly rises with [Fe/H] until it reaches a
maximum and then declines again for the most metal-rich stars (RGB-a according
to \cite{pancino:p00}). This nicely confirms a previous finding by
\cite{pancino:p02} and \cite{pancino:o03}. If the metal-rich stars have evolved
within the cluster in a process of self-enrichment, the only way to lower their
$\alpha$-enhancement would be SNe type Ia intervention. No simple explanation
is provided for the rise of [Ca/Fe] at low [Fe/H], although a series of star
formation bursts should be the likely cause.

%

\end{document}